\documentclass[preprint,showpacs,preprintnumbers,amsmath,amssymb]{revtex4}
\usepackage{graphicx}
\usepackage{dcolumn}
\usepackage{bm}

\begin{document}
\title{Entropic Forces and Bendability of short DNA helices}

\author{Marco Zoli}

\affiliation{School of Science and Technology \\  University of Camerino, I-62032 Camerino, Italy \\ marco.zoli@unicam.it}

\date{\today}

\begin{abstract}
The flexibility of short DNA fragments is studied by a Hamiltonian model which treats the inter-strand and intra-strand forces at the level of the base pair. The elastic response of a set of homogeneous helices to externally applied forces is obtained by computing the average bending angles between adjacent base pairs along the molecule axis. The ensemble averages are performed over a room temperature equilibrium distribution of base pair separations and bending fluctuations. The analysis of the end-to-end distances and persistence lengths shows that even short sequences with less than $100$ base pairs maintain a significant bendability ascribed to thermal fluctuational effects and kinks with large bending angles. The discrepancies between the outcomes of the discrete model and those of the worm-like-chain model are examined pointing out the inadequacy of the latter on short length scales.
\end{abstract}


\maketitle

\section*{I. Introduction}

Thermal fluctuations constantly deform the DNA molecular bonds crumpling the helix into a random coil configuration whose end-to-end distance ($L_{e-e}$) is much shorter than the overall contour length ($L$). Accordingly, thermal bending fluctuations are key to the DNA biological functioning as they bring together enhancers (regulatory DNA sequences to which activator proteins bind) \textit{and} gene promoter regions (stretches of genome where DNA transcription factors bind) that could be even thousands of nucleotide pairs away along the molecule axis. 
Such long range effects \cite{yera1} allow transcription processes which  control gene expressions and ultimately cell differentiation \cite{bird}.
While in the cell's aqueous environment the DNA coiled configuration is associated to a large conformational entropy, a transition to a stretched elastic regime \cite{odi,marko95} may occur in vivo under the action of binding proteins which significantly stretch the DNA polymer chain \cite{gelles,marko98}. 

Bending flexibility and stretching properties of single DNA molecules have been quantitatively analyzed following the development of micro-manipulation techniques and force transducers in the pico-Newton regime  \cite{chu,busta92}. Forces in such range permit to straighten bends and kinks in the helix due to the buffeting of the thermal bath. In fact, at the nano-scale, the room temperature thermal energy per nano-meter is, $k_B T / nm \sim 4 \, pN$. 

In general, the strength of the applied force sets the length scale in the single molecule elastic response with larger forces sampling shorter scales, down to the covalent bond distance between adjacent nucleotides along the strands \cite{raposo}.

Groundbreaking experiments on $\lambda$-phage DNA \cite{cluzel,busta94,busta96} have revealed a force versus extension pattern with distinct regimes.
In the limit of weak applied forces, the chain extension is lower than $L$ and DNA behaves as a linear spring whose force constant is inversely proportional both to the persistence length ($l_p$) and to $L$. By increasing the force in the intermediate regime, more work is done against the random fluctuations of the thermal bath and the DNA behavior becomes progressively non-linear. 
 
For larger forces, $L_{e-e}$ gets of order of $L$, chemical changes intervene in the molecule backbone and the chain elastic response should include the enthalpic compliance through a finite stretch modulus. 
At $\sim 65 \, pN$ the molecule elongates abruptly from a length of $\sim 1.1 \,L$ to $\sim 1.7 \,L$ in a very narrow force range displaying a plateau in the force-extension relation which is the peculiar feature of a cooperative over-stretching transition. Importantly, the latter is observed at $\sim 65 \, pN$ if the experimental set-up allows the $\lambda$-phage molecule to untwist while being stretched. On the other hand, once both ends of the two complementary strands are anchored such as the molecule is torsionally constrained  \cite{strick98,balter}, the sudden elongation is shifted at higher forces, namely it shows up  above $\sim 100 \, pN$ \cite{mameren}. 
While it is still debated whether the mechanically over-stretched $\lambda$-molecules transform into a structurally different S-DNA or simply melt in two separate strands \cite{rief,rouzina,yan12,palmeri,bianco14}, it is recognized that in the small and intermediate regime, up to applied forces of $\sim 10 \, pN$, the extension data are well described by Worm-like-Chain (WLC) elastic models \cite{busta00}.  Importantly, it is pointed out that $\lambda$-molecules are sufficiently long, $L=\, 16.4 \, \mu m$, to fulfill the key assumption for the application of the WLC approach that is, $l_p \ll L$, with $l_p$ being typically of $ \sim 50\, nm$ in DNA although intrinsic values of  $ \sim 40\, nm$ have also been reported \cite{block}.

Instead, for short fragments with $L$ in the range of only a few tens of base pairs (\textit{bps}), 
recent advances in experimental methods have permitted to characterize physical properties which challenge the view of the WLC model.

DNA cyclization measurements using both the ligase protein assays \cite{widom} and, more recently, single-molecule fluorescence resonance energy transfer (FRET) assay \cite{vafa,kim13} have delivered $J$- factors which are several orders of magnitude larger than predicted by the conventional WLC model \cite{shimada}.  Small-angle x-ray scattering (SAXS) measurements of mean and variance of $L_{e-e}$ have been explained in terms of soft stretching modes which cooperatively involve \textit{bps} over two helix turns \cite{fenn}. These results have been corroborated by molecular dynamics simulations which find a softening by almost one order of magnitude in the stretching modulus of a 56-mer mainly ascribed to end effects  \cite{gole}. Also combined FRET measurements of $L_{e-e}$ and SAXS measurements of the radius of gyration
for a set of sequences with $15 - 89$ \textit{bps} \cite{archer} point to a remarkable chain flexibility, possibly due to the occurrence of single base pair breaking, and are consistent with a substantially lower $l_p$ than the typical value. 

While extensions of the WLC model have also been proposed  to interpret the controversial results of all these 
measurements \cite{mastro,menon,eve,volo13,mazur,kim14}, there is growing consensus that short DNA fragments may indeed display large bending fluctuations and persistence lengths which are significantly smaller than those traditionally estimated for long molecules.  
Thus, the overall picture emerging from a body of experimental and theoretical work is that the DNA elastic properties vary with the molecule size and that an intrinsic flexibility emerges at those length scales relevant to biological processes such as regulation of protein binding and formation of nucleosomes whose basic unit is indeed a fragment of $147$ \textit{bps}. 

Certainly a comprehension of the DNA phenomenology at such short scales requires a characterization of the model based on the intermolecular forces which are instead coarse grained in the standard WLC approach. This in fact considers the DNA molecule as a flexible chain, inextensible along its contour, whereby the orientational correlation function between distant segments decays exponentially over the characteristic $l_p^{WLC}$ which amounts to a length of $\sim 150$ \textit{bps} along the stack. Here we propose a method that treats the DNA molecule at the level of the base pair and calculates the chain bending deformations by a mesoscopic Hamiltonian which includes the hydrogen bonds between complementary strands and the stacking interactions along the molecule backbone. The latter bear dependence on both the helix twisting and the bending angles between adjacent \textit{bps}. Performing ensemble averages over equilibrium distributions of base pair fluctuations we obtain quantitative estimates for experimentally accessible quantities such as $L_{e-e}$ and  $l_p$ for a set of DNA molecules in terms of the input parameters of the mesoscopic potential. 
While method and  model are general and can be applied to any heterogeneous sequence, we focus here on a set of homogeneous chains highlighting the effects of their length.

The model contains a tunable external force which perturbs the molecule by coupling to the specific sites along the stack. Assuming a initial fully stretched configuration, $L_{e-e}=\,L$, we introduce a small force which induces the elastic response of the molecule dominated by random bending due to thermal fluctuations. Accordingly $L_{e-e}$ gets smaller than $L$ in the weak force regime. This occurs as long as the applied force becomes sufficiently strong to overcome the contraction due to the intrinsic entropic term. At this stage the molecule begins to stretch,  $L_{e-e}$ grows as a function of the external force and eventually tends to the contour length value. 
It is shown that the length of the chain critically affects the response to the external field hence, the transition between entropic regime and intermediate force regime in which the molecule stretches. Thus, simulating various force profiles acting at the base pair level, we establish quantitative relations between applied strengths and molecule macroscopic properties as a function of the chain length.
In this context, the reduced $l_p$ values found for short molecules are interpreted as a consequence of the contractile forces and are consistent with the presence of large bending fluctuations between adjacent \textit{bps}. The results of our discrete method are compared with the predictions of the WLC model while a non strict comparison is given with the experimental data available for a short sequence in the range of those here considered. The large forces regime in which rise distance and DNA structure are altered is not investigated in this work.

In Section II, we describe the geometrical model for the DNA molecule whose axis is subject to the action of the intrinsic contractile force and of the applied forces. The mesoscopic Hamiltonian  is proposed in Section III while the method to compute the average bending angles between the base pair planes is explained in Section IV.
The general relations between end-to-end distance and persistence length are reported in Section V. The obtained results are discussed in Section VI and some final remarks are made in Section VII.

\section*{II. Helix with Entropic Forces}

\begin{figure}
\includegraphics[height=8.0cm,width=8.0cm,angle=-90]{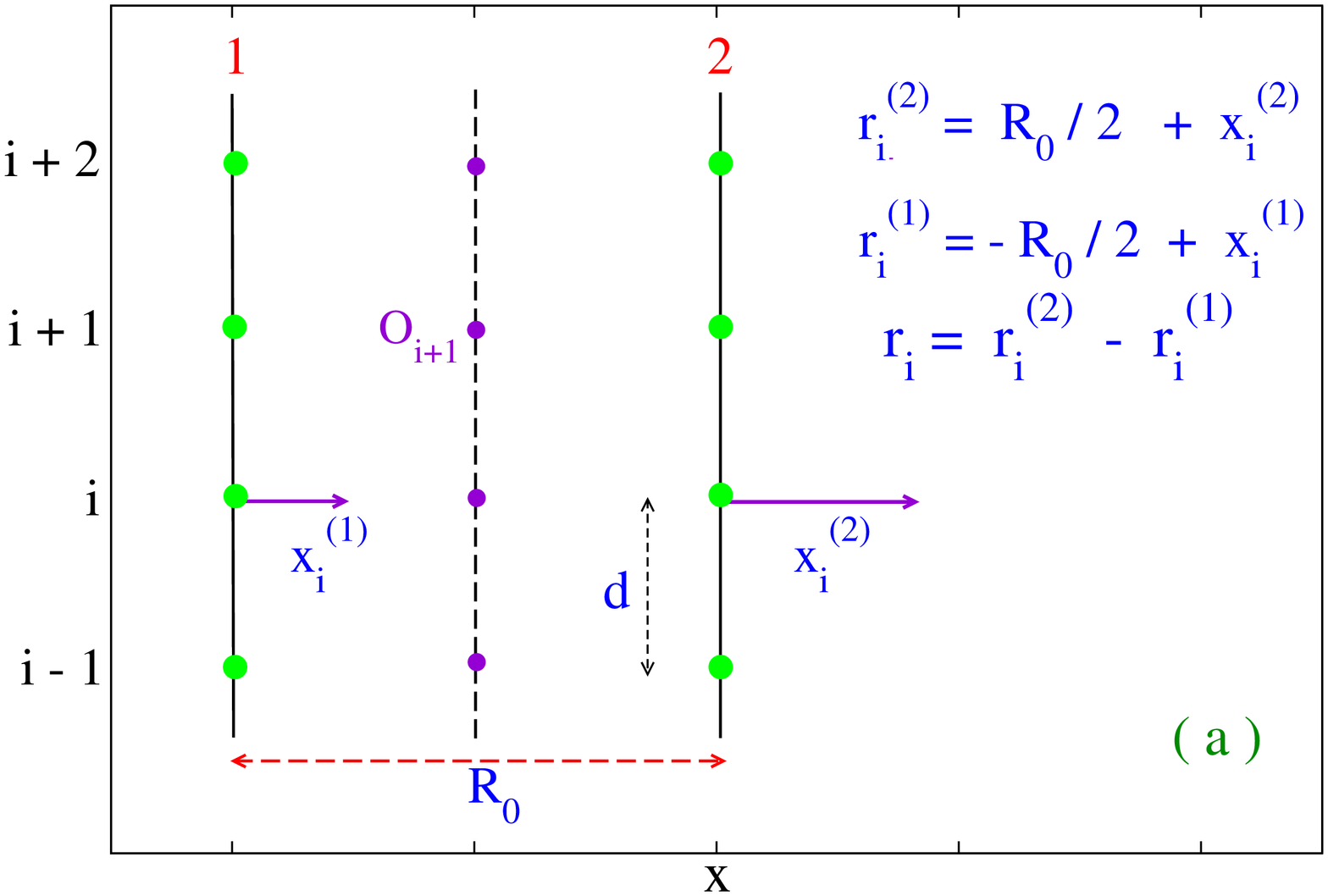}
\includegraphics[height=8.0cm,width=8.0cm,angle=-90]{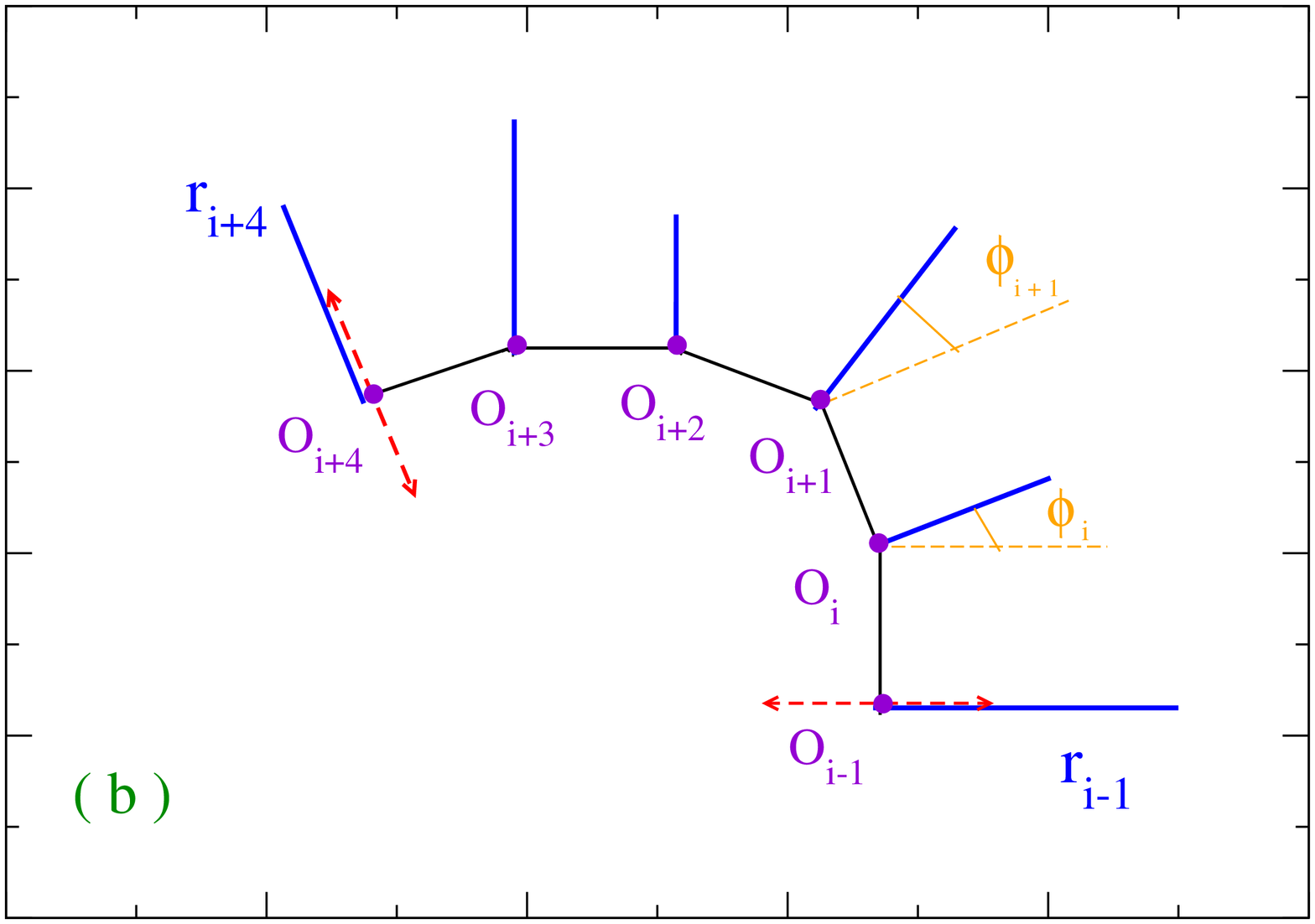}
\caption{\label{fig:1}(Color online)  
(a) Schematic of the model for $N$ base pairs in a ladder representation.  $N$ bases (green dots) are stacked along each of the two complementary strands. $R_0$ is the equilibrium inter-strand distance and $d$ is the rise distance. $x_{i}^{(1,2)}$ are the fluctuations of the $i-th$ base pair mates with respect to the equilibrium.
$r_{i}$ is the relative distance between the two mates. The $O_i$'s lie along the central helical axis. {}  (b)  The helix axis is a chain of $N - 1$ segments of length $d$ connecting the $O_i$'s which are pinned to the sheet plane, hence the helix axis is planar. The (red) long-dashed lines, drawn at the $O_{i-1}$ and $O_{i+4}$ sites, denote the helix diameter $R_0$.
The $r_i$'s depart from the $O_i$'s, have variable amplitudes and are parallel to $R_0$ at their respective sites. $\phi_{i}$ is the (variable) bending angle between adjacent base pair vectors. The $\phi_{i}$'s are measured from the (orange) short-dashed lines which are parallel to the adjacent (preceding) $r_{i-1}$'s along the chain.
}
\end{figure}

The basic representation for a double stranded chain is based on a ladder model as shown in Fig.~\ref{fig:1}(a). 
The two mates of the $i-th$ base pair can fluctuate around their equilibrium positions represented by the green dots lying along the two complementary strands. 
The vibrations of the two bases along the stack are much smaller than the transverse vibrations $x_{i}^{(1,2)}$, i.e. the model is at this stage one-dimensional. $x_{i}^{(1)}$ and $x_{i}^{(2)}$ may be in-phase (as depicted) or out-of-phase. In general, also their amplitudes may differ. 
$R_0 \sim \,20 $\AA {} is the average helix diameter and  $d \sim \, 3$\AA {} is the average rise distance. 
With respect to the central helical axis (that is kept fixed), we build the vectors $r_{i}^{(1)}=\, -R_0 / 2 + x_{i}^{(1)}$ and $r_{i}^{(2)}=\, R_0 /2 + x_{i}^{(2)}$ and define the relative distance $r_{i}=\, r_{i}^{(2)} - r_{i}^{(1)}$. We observe that: \textit{i)} also in-phase vibrations of different amplitudes may contribute to $r_{i}$ shifting the base pair out of the stack.  \textit{ii)} $r_{i}$ may shrink with respect to $R_0$ but too large contractions are prevented by the strands electrostatic repulsion.

Next, we go beyond the ladder model and admit that the radial displacements $r_{i}$ and $r_{i-1}$ along the stack are bent by the angle $\phi_i$, shown in Fig.~\ref{fig:1}(b), which is an integration variable of our model. In general, the $r_{i}$'s are not constrained to the sheet plane: adjacent $r_{i-1}$ and $r_{i}$  along the stack are in fact twisted by an angle $\theta_i$ (not drawn) lying on a plane normal to the sheet \cite{io11}. For the $i-th$ base pair, the torsional angle is given by, $\theta_i =\, (i - 1)\theta  + \theta_S$,  with $\theta=\, 2\pi / h$  and $h$ is the helical repeat, i.e., the number of \textit{bps} per helix turn. The standard value measured for instance in covalently closed DNA in solution is \, $\,h \sim \,10$ \cite{wang}. More generally, one may release the torsional constraint and consider $h$ as a variable to be determined by free energy minimization \cite{io14} for any value of the applied force. This would describe the twist-stretch coupling effect for short chains.
As $\theta_S$ is the twist of the first base pair along the stack, a sum over $\theta_S$'s is included in the partition function to allow for a distribution of possible \textit{bps} orientations.

Thus our model  is essentially made by $N-1$ segments connecting the $O_i$'s which are arranged as beads along the central helical axis. For any $O_i$, there is a base pair distance $r_{i}$ which independently fluctuates on the plane (normal to the sheet) containing the average helix diameter, drawn by the (red) long-dashed line in Fig.~\ref{fig:1}(b). Furthermore, adjacent $r_{i}$'s along the stack are twisted by a variable $\theta_i$ and bent by a variable $\phi_i$ accounting for the rotational degrees of freedom.

\begin{figure}
\includegraphics[height=8.0cm,width=8.0cm,angle=-90]{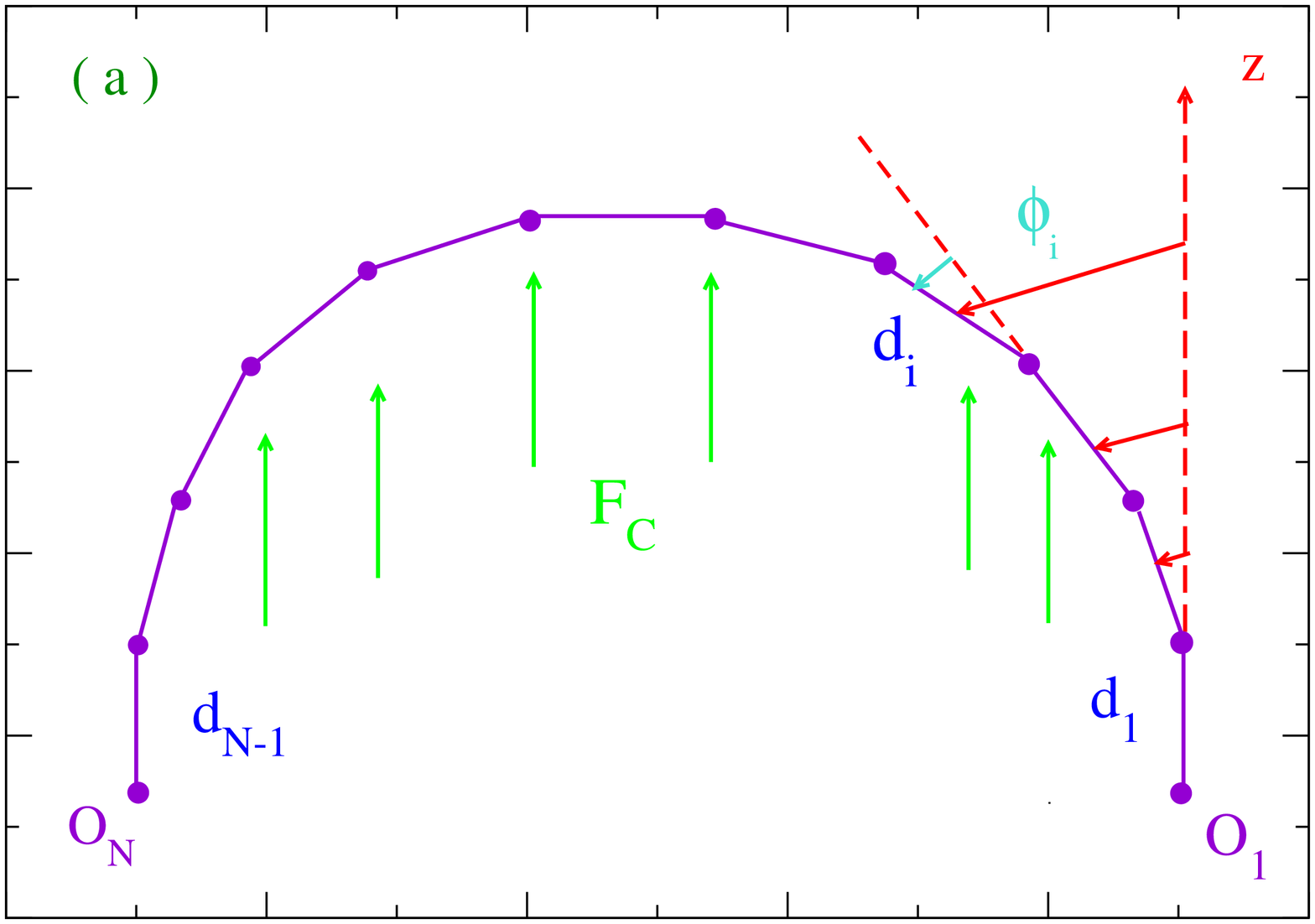}
\includegraphics[height=8.0cm,width=8.0cm,angle=-90]{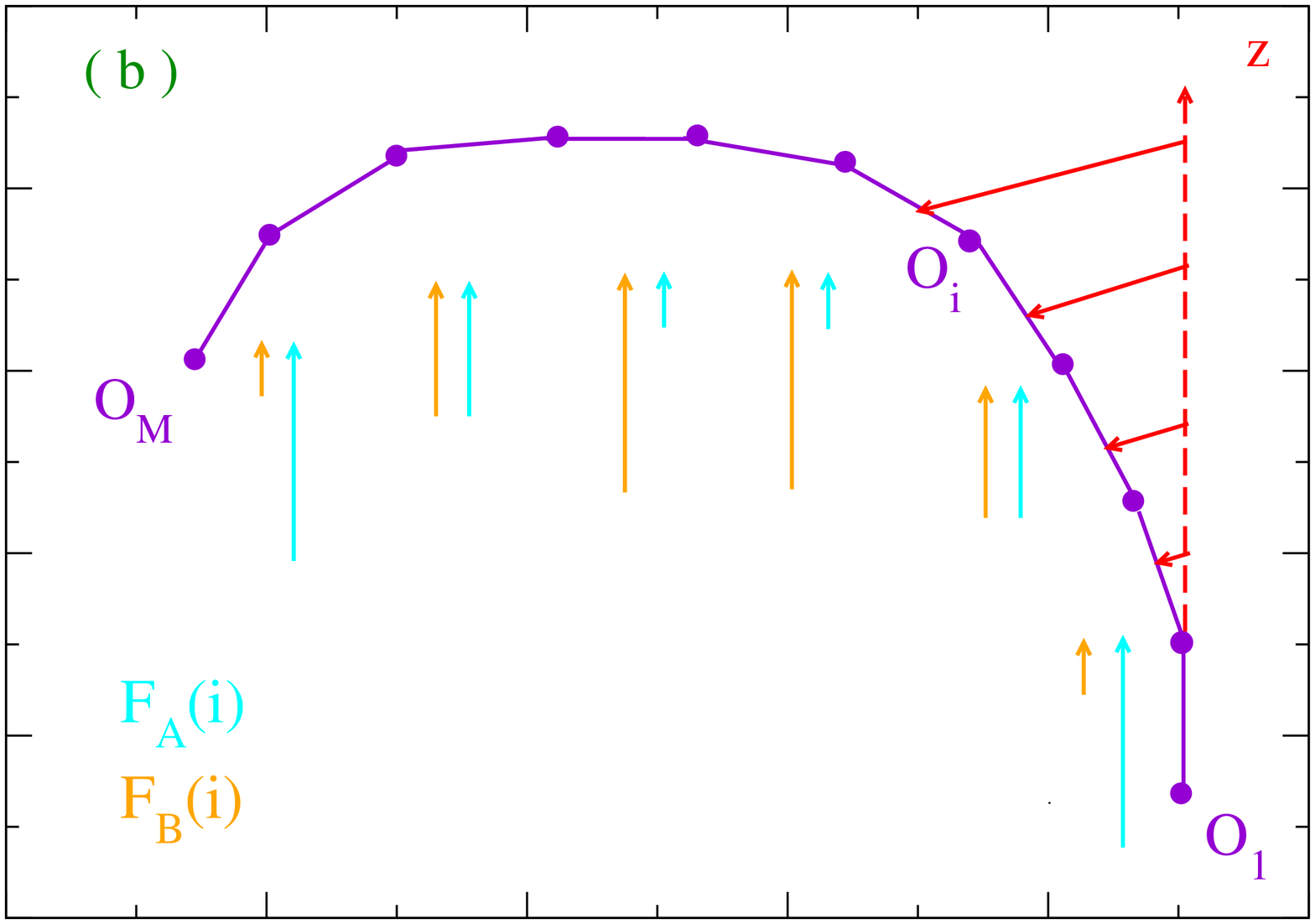}
\caption{\label{fig:2}(Color online)  
(a)  {} Chain with $N$ beads and $N-1$ bonds with a constant (site independent) force acting along the $z$ direction. The $O_i$'s are arranged on the helical axis as in Fig.~\ref{fig:1}. $\phi_{i}$ is the bending angle between the segments $d_i$ and $d_{i-1}$ whose constant modulus is $d$. The base pair distances $r_i$'s are not drawn here. The end bond, $d_{N-1}$, is bent by an angle $\pi $ with respect to the first bond, $d_1$.  (b) {} A shorter chain, $M < N$, under the action of site dependent forces. The maximum angle formed by the end bonds, $d_{M-1}$ and $d_1$ is smaller than $\pi $.}
\end{figure}

Note that the $O_i$'s are always pinned to the sheet plane despite the presence of the twist degree of freedom. This property ensures that, once the short fragments form a loop,  the writhe of that circular molecule is zero \cite{bates,lionb,irob}. Accordingly the model can be extended to study also the $J$- factors of short fragments by imposing suitable boundary conditions. While the cyclization probabilities are expected to decrease by shortening the molecule length, the entropic forces may substantially bend even short linear chains. As represented in Fig.~\ref{fig:2}(a) for a chain with $N$ beads, the first bond is assumed to be aligned along a reference z-axis, the successive bonds connecting the beads are bent by  $\phi_i$'s and the last ($N-1$) bond makes an angle $\pi $ with respect to the z-axis. The applied force
$F_C$ is constant.
In general, in very short chains, the overall bending angle may be smaller than $\pi $ as depicted in Fig.~\ref{fig:2}(b). Moreover, the external forces  may not be constant throughout the chain. For instance, by $F_A(i)$ we simulate a force which is maximum at the chain ends whereas $F_B(i)$) is maximum at the center of the molecule. These different scenarios are compared in the calculations presented in Section V.

\section*{III. Hamiltonian Model}

Formally our physical picture, for an open ends molecule with finite helical radius and base pairs with reduced mass $\mu$, is described by the Hamiltonian:

\begin{eqnarray}
& &H =\, H_a[r_1] + \sum_{i=2}^{N} H_b[r_i, r_{i-1}] \, , \nonumber
\\
& &H_a[r_1] =\, \frac{\mu}{2} \dot{r}_1^2 + V_{1}[r_1] \, , \nonumber
\\
& &H_b[r_i, r_{i-1}]= \,  \frac{\mu}{2} \dot{r}_i^2 + V_{1}[r_i] -\textbf{ F}_J(i-1) \cdot \textbf{d}_{i-1} + V_{2}[ r_i, r_{i-1}, \phi_i, \theta_i]   \, , \nonumber
\\
& &V_{1}[r_i]=\, V_{M}[r_i] + V_{Sol}[r_i] \, , \nonumber
\\
& &V_{M}[r_i]=\, D_i \bigl[\exp(-b_i (|r_i| - R_0)) - 1 \bigr]^2  \, , \nonumber
\\
& &V_{Sol}[r_i]=\, - D_i f_s \bigl(\tanh((|r_i| - R_0)/ l_s) - 1 \bigr) \, , \nonumber
\\
& &V_{2}[ r_i, r_{i-1}, \phi_i, \theta_i]=\, K_S \cdot \bigl(1 + G_{i, i-1}\bigr) \cdot (r_i - r_{i-1})^2  \, , \nonumber
\\
& &G_{i, i-1}= \, \rho_{i, i-1}\exp\bigl[-\alpha_{i, i-1}(|r_i| + |r_{i-1}| - 2R_0)\bigr]  \, , \nonumber
\\ 
& &\textbf{F}_J(i-1) \cdot \textbf{d}_{i-1}=\, F_J(i-1) d \cos\biggl( \sum_{k=1}^{i-1}\phi_k \biggr) \, . \nonumber
\\
\label{eq:02}
\end{eqnarray}

Every base pair in the chain (except the two end sites) interacts with its two adjacent neighbors. As only the first site, $i=\,1$, lacks the preceding base pair along the chain,  its kinetic term and one particle potential have been treated separately by defining $H_a[r_1]$. 
Moreover, the first site is coupled to the second one via the $i=\,2$ term in $H_b[r_i, r_{i-1}]$.

The one particle potential, $V_{1}[r_i]$, is made of two contributions: 

\textit{(1)} the Morse potential, $V_{M}[r_i]$, models the hydrogen bond between the two mates of the $i$- base pair: $D_i$ is the pair dissociation energy and $b_i$ determines the potential range. The relative base distances are measured with respect to the helix diameter that sets the zero for the potential. 

Fluctuations in the base pair separations may reduce the distance $r_i$ between complementary strands to values smaller than $R_0$, a case also contemplated at some sites in Fig.~\ref{fig:1}(b).
However, such reduction is limited by the hard core electrostatic repulsion due to the negatively charged phosphate groups. To comply with this physical requirement, the numerical code discards $r_i$ such that, $|r_i| - R_0 < - \ln 2 / b_i$ which would deliver a repulsive energy larger than $D_i$. 

In heterogeneous chains, Adenine-Thymine \textit{bps} can be broken more easily by thermal fluctuations and undergo larger stretching vibrations than Guanine-Cytosine \textit{bps} \cite{campa,kalos}.  Thus, in general, the Morse parameter should satisfy the following inequalities: $D_{AT} < D_{GC}$ and $b_{AT} < b_{GC}$  although substantial variations regarding their values have been reported \cite{zdrav,singh,weber09,weber15}.  While the mesoscopic model in  Eq.~(\ref{eq:02}) can be applied to study sequence specificities \cite{io14a}, the present investigations essentially focuses on  length scale effects. Accordingly, the molecules are assumed as homogeneous. The input parameters $D_i$ and $b_i$ are set to yield a free energy per base pair in line with the experimental data \cite{kame06,metz11}, i.e., $D_i=\,60 \,meV$ and $b_i=\,5\,\AA^{-1}$.

\textit{(2)} The solvent potential, $V_{Sol}[r_i]$, accounts for the fact that DNA is immersed in water and its physical properties depend on the salt concentration \cite{owc,zanc,singh15}. Then, the molecules stability can be empirically related to the $f_s$ parameter \cite{druk}. As a main effect, the solvent potential enhances by $f_s D_i$ (with respect to the Morse plateau) the height of the energy barrier above which the base pair dissociates. Thus, the full one particle potential, $V_{M}[r_i] + V_{Sol}[r_i]$, shows a hump whose width is tuned by $l_s$. This length defines the range within which $V_{Sol}$ is superimposed to the plateau of the Morse potential.   
See e.g., refs.\cite{io11,io12} for a broader discussion of the solvent effects.

The constant force ${F}_C$ in Fig.~\ref{fig:2}(a) and the site dependent forces \, ${F}_J(i)$ \, ($J=\,A, B$) in Fig.~\ref{fig:2}(b), are coupled to the intra-strand bonds whose modulus, i.e. the rise distance, is taken constant as the forces hereafter discussed are in the weak to intermediate range.   $F_C$ and ${F}_J(i)$ act along the direction of the first bond, i.e., $\phi_1=\,0$. In the weak forces regime, the molecule response is dominated by the fluctuations in the bending angles $\phi_i$  between adjacent chain segments which are ultimately responsible for the molecule end-to-end contraction depicted in Figs.~\ref{fig:2}. By increasing the intensity of the applied forces, the chain segments straighten and the molecule end-to-end distance returns to grow. 
Note that the overall angle formed by $d_i$ with respect to the $z$-axis is written as a sum over the bending angles formed by the preceding segments. For any $d_i$, we compute an average $< \phi_i >$ over an ensemble of base pair fluctuations which are coupled by the stacking potential.
This establishes the chain correlations which determine the persistence length in this model. 

The stacking interactions are modeled by a non-linear two particles potential, $V_{2}[ r_i, r_{i-1}, \phi_i, \theta_i ]$, in which the square distance between adjacent $r_i$ and $r_{i-1}$ in Eq.~(\ref{eq:02})  depends both on the twist and on the bending angles.
The basic form of this potential was originally proposed, in the context of a ladder DNA model \cite{pey2}, to account for those cooperative effects which propagate along the molecule stack forming thermal bubbles and DNA denaturation at high temperature.  

The underlying idea is that, whenever $r_{i}  - R_0 \gg \alpha_{i, i-1}^{-1}$, the $i-th$ hydrogen bond is broken and the stacking coupling drops from \, $\sim K_S \cdot (1 + \rho_{i, i-1})$ to $\sim K_S$:  this also favors the breaking of the adjacent base pair and the consequent opening of local bubbles \cite{benham,bonnet,ares,rapti,palmeri13}. Then, small $\alpha_{i, i-1}$  indicate that  large fluctuations are required to unstack a base pair and produce such a reduction in the stacking. The values $K_S=\,10 \, meV \AA^{-2}$, $\alpha_{i, i-1}=\,2.5 \AA^{-1}$ and $\rho_{i, i-1}=\,1$ are taken hereafter.

Beyond these nonlinear properties, $V_{2}$ also incorporates the angular degrees of freedom of the helicoidal model.
In this regard, our stacking potential is more complex than the usual elastic terms assumed in WLC models \cite{volo2000}.
While the specific $V_{2}$ chosen in Eq.~(\ref{eq:02}) has robust physical motivation, it is remarked that different potentials may be taken e.g., with the purpose to ensure the finiteness of the intra-strand stacking also for large inter-strand separation \cite{joy09}. This requirement is fulfilled by truncating the phase space available to the base pair separations as described in the next Section. 
  
Importantly, our computational technique has the advantage to tackle the divergence of the partition function for the Hamiltonian in Eq.~(\ref{eq:02}), encountered e.g., in transfer integral techniques \cite{zhang}. Such problem arises from the fact that the one-particle potential is bounded for $r_i \rightarrow \infty$. Then, if all $r_i$'s are equal (translational mode) and infinitely large, the two-particles potential vanishes while $H$ remains finite hence the partition function diverges. This zero mode cannot be removed via standard techniques \cite{schulman} due to the lack of translational invariance caused by the on-site potential.

\section*{IV. Method}

The idea underlying our method is that the base pair separations can be mapped onto the time axis, $r_i \rightarrow |r_i(\tau)|$,  so that the distance between the base pair mates is a trajectory depending on the imaginary time  $\tau=\,it$,  with $t$ being the real time for the evolution amplitude within the time interval, $t_b - t_a$. 

The theoretical grounds of the method lie in the analytic continuation of the quantum mechanical partition function to the imaginary time axis which, in general, permits to obtain the quantum statistical partition function \cite{fehi}. Accordingly $\tau$ varies in a range $\tau_b - \tau_a$ whose amplitude is set by the inverse temperature $\beta$
and the partition function is written as an integral over closed trajectories, $(\,r_i(0)=\, r_i(\beta) \,)$,  running along the $\tau$-axis. 
While the imaginary time formalism is widely used in semi-classical methods for the solution of quantum statistical problems \cite{jack}, 
it can also be extended to treat the room temperature classical ensemble of DNA molecules as extensively described in refs.\cite{io09}. The main features of the method are hereafter outlined.

As a consequence of the $\tau$-closure condition,  the $r_i(\tau)$ are expanded in Fourier series around $R_0$:

\begin{eqnarray}
& &r_i(\tau)=\, R_0 + \sum_{m=1}^{\infty}\Bigl[(a_m)_i \cos(\omega_m \tau ) + (b_m)_i \sin(\omega_m \tau ) \Bigr] \, \nonumber
\\
& &\omega_m =\, \frac{2 m \pi}{\beta} \,.
\label{eq:03}
\end{eqnarray}

Accordingly the integration measure $\oint {D}r_i$  is defined over the space of the Fourier coefficients:

\begin{eqnarray}
& &\oint {D}r_{i} \equiv  \prod_{m=1}^{\infty}\Bigl( \frac{m \pi}{\lambda_{cl}} \Bigr)^2 \int_{-\Lambda_T}^{\Lambda_T} d(a_m)_i \int_{-\Lambda_T}^{\Lambda_T} d(b_m)_i \, , \, \label{eq:04}
\end{eqnarray}

where $\lambda_{cl}$ is the classical thermal wavelength and $\Lambda_T$ is the temperature dependent cutoff. 

The expansion in Eq.~(\ref{eq:03})  generates a large ensemble of trajectories for any base pair. Say \, $2N_p + 1$ \, the number of integration points for each Fourier coefficient  in Eq.~(\ref{eq:04}). Then, for a single Fourier component in Eq.~(\ref{eq:03}), the computation includes \, $(2N_p + 1)^2 \cdot N_\tau $ \, base pair states where $N_\tau $ is the number of integration points over the time axis. The total number of base pair states sets the size of the base pair ensemble in the phase space. Such number is increased until, taking  $N_p=\,300$ and $N_\tau=\,100$,
numerical convergence in the partition function is achieved. This ensures that the physical properties of the chain are computed over an equilibrium distribution of states. The latter have to fulfill the physical requirements described in Section III. Thus our numerical program selects, at any $T$, an ensemble of base pair trajectories which are \textit{both} consistent with the model potential constraints \textit{and} in accordance with the second law of thermodynamics \cite{io11}.

Moreover, the measure in Eq.~(\ref{eq:04}) normalizes the kinetic term in the action \cite{io05}, i.e.:

\begin{eqnarray}
\oint {D}r_i \exp\Bigl[- \int_0^\beta d\tau {\mu \over 2}\dot{r}_i(\tau)^2  \Bigr] = \,1 \, .
\label{eq:11} \,
\end{eqnarray}

This condition holds for any $\mu$ hence, the free energy does not depend on $\mu$ as expected for a classical system.
Peculiar to the method is also the fact that, through Eqs.~(\ref{eq:04}),~(\ref{eq:11}), we precisely determine the cutoffs $\Lambda_T$ in the multiple integrations over the Fourier coefficients  \cite{io11a} thus avoiding those indeterminacies encountered in the transfer integral methods. Moreover, through the $T$-dependent cutoffs, the present method builds base pair amplitudes which are growing function of temperature in agreement with general expectations  \cite{zocchi03}.

Using Eq.~(\ref{eq:04}) and consistently with the notation for the Hamiltonian  in Eq.~(\ref{eq:02}), we can write the partition function $Z_N$ for the ensemble of molecules with $N$ \textit{bps}:

\begin{eqnarray}
& &Z_N=\, \oint Dr_{1} \exp \bigl[- A_a[r_1] \bigr]   \prod_{i=2}^{N} \sum_{\theta_S} \int_{- \phi_{inf} }^{\phi_{sup} } d \phi_i  \oint Dr_{i}  \exp \bigl[- A_b [r_i, r_{i-1}] \bigr] \, , \nonumber
\\
& &A_a[r_1]= \,  \int_{0}^{\beta} d\tau H_a[r_1(\tau)] \, , \nonumber
\\
& &A_b[r_i, r_{i-1}]= \,  \int_{0}^{\beta} d\tau H_b[r_i(\tau), r_{i-1}(\tau)] \, ,
\label{eq:05}
\end{eqnarray}

where the action $A_b[r_i, r_{i-1}]$ depends: \textit{(1)} on the Fourier coefficients 
$\{ (a_m)_i, \, (b_m)_i\}$ and $\{ (a_n)_{i-1}, \, (b_n)_{i-1}\}$
of the $i$ and $i-1$ base pair path amplitudes respectively; \textit{(2)} on the angles $\theta_i$ and $\phi_i$. Kinks with even large bending angles are included in Eq.~(\ref{eq:05}) via the angular cutoffs  ($\phi_{inf}$, $\phi_{sup}$) as discussed below.
Thus, the two particle stacking potential brings about a mixing of the Fourier components of adjacent \textit{bps} which is mainly responsible for the chain stiffness.

This is evaluated by computation of the ensemble average for the bending angles, over an equilibrium distribution of base pair states, defined by:

\begin{eqnarray}
& &< \phi_j >_{(j \geq 2)} =\,  \frac{\sum_{\theta_S} \int_{ \phi _{inf}}^{\phi _{sup}} d \phi_j \cdot \phi_j \oint Dr_{j} \exp \bigl[- A_b [r_j, r_{j-1}]  \bigr]}{\sum_{\theta_S} \int_{\phi _{inf} }^{\phi _{sup} } d \phi_j  \oint Dr_{j} \exp \bigl[- A_b [r_j, r_{j-1}]  \bigr]} \, , \nonumber
\\
& &\phi _{inf}=\,< \phi_{j-1} > - \phi_M/2 \, , \nonumber
\\
& &\phi _{sup}=\,< \phi_{j-1} > + \phi_M/2 \, , 
\label{eq:06}
\end{eqnarray}

As depicted in Figs.~\ref{fig:2}, the first segment is aligned along the z-axis, i.e., $< \phi_{1} >=\, 0$. For each $j$-segment, the average angle is computed by summing over fluctuations in a range of width $\phi_M =\,\pi /2$ centered on the average angle for the preceding segment, that is $< \phi_{j-1} >$. It follows that, for any  distant $k-$ and $j-$th \textit{bps} of the chain, the relative average bending is calculated as

\begin{eqnarray}
< \phi_{j,k} > =\, \sum_{l=j+1}^{k} < \phi_l >  \,.
\label{eq:06a}
\end{eqnarray}

Eqs.~(\ref{eq:06}), ~(\ref{eq:06a}) form the bridge between the  Hamiltonian description in terms of the intermolecular forces and the measurable macroscopic properties of the chain.
Finally, it is remarked that:

\textit{a)} although all bonds drawn in Figs.~\ref{fig:2} rotate counterclockwise,  also clockwise angles between adjacent bonds are possible according to Eq.~(\ref{eq:06}).

\textit{b)} The ensemble averages are performed by integrating both over a range of bending angles and, importantly, over a distribution of $r_i$'s. However, from Eqs.~(\ref{eq:04}) and ~(\ref{eq:06}), it appears that our method deals with base pair separations and angular degrees of freedom on a different footing.
While the $r_i$'s are Fourier expanded and a multiple integration with $T$-dependent cutoff is performed, the angular variables are integrated out in a conventional way.   In fact, at the present stage there is not enough knowledge regarding the temperature dependence of the bending angles to justify a computationally time consuming approach such to incorporate also a $T$-dependent cutoff on the bending fluctuations. 

\textit{c)} The $< \phi_{j} >$ also depend on the twist angles distribution through the sum over $\theta_S$. While in the following calculations the helical repeat that is, the twist between adjacent \textit{bps}, is set to a constant value, the method can also account for the more general case in which $h$ is a variable. This holds whenever the helix (un)twisting is coupled to the stretching/bending of the molecular axis as in the experimental configuration discussed in the Introduction.

\section*{V. End-to-End Distance and Persistence Length}

To analyze the flexibility of short DNA fragments we set out to calculate the mean square end-to-end distance for a chain with $N-1$ segments shown in Figs.~\ref{fig:2}. A picture is adopted in which the contour length is \, $L=\,(N-1)d$ \, whereas one may more precisely take \, $L=\,\sum_{i=2}^{N}|<{r}_i> - <{r}_{i-1}>|$, with the base pair distances given in Fig.~\ref{fig:1}(b). This latter choice is however not essential to our purpose of studying the interplay between applied forces and bendability as a function of $N$. Furthermore, the contour length is known to depend essentially on the stretching rigidity of the chain \cite{tan} which is taken constant in this work. 
Then we assume that the base pair separations only contribute through the ensemble averages in Eq.~(\ref{eq:06}) and consistently 
define the end-to-end vector, $\textbf{D}=\,\sum_{i=1}^{N-1}\textbf{d}_i$.  Then \, $L_{e-e}^2\equiv \, < D^2 >$ reads:

\begin{eqnarray}
& & L_{e-e}^2  =\,  \sum_{i=1}^{N-1} \sum_{j=1}^{N-1} < \textbf{d}_i \cdot \textbf{d}_j >\, \nonumber
\\
& &< \textbf{d}_i \cdot \textbf{d}_j >  =\, d^2 < \cos \phi_{i,j} > \, 
\label{eq:07}
\end{eqnarray}

where the average angles are computed as described in the previous Section. 

In general the correlation between distant segments, depending on the sequence specificities and on those intermolecular forces which shape the polymer structure \cite{trifo}, are lost because of thermal fluctuations or specific interactions with the solvent molecules \cite{save}.
As for homogeneous sequences the chain directionality can be neglected, two segments are correlated over the distance spanned by $2k$ consecutive segments such that $< \textbf{d}_i \cdot \textbf{d}_{i\pm k} >\neq 0$. 
Such correlation distance is quantitatively determined by the site dependent persistence length  \cite{soder} :

\begin{eqnarray}
l_p(i) =\,  d \biggl(1 +  \sum_{j > i}^{N-1} < \cos \phi_{i,j} > \biggr)\, 
\label{eq:08}
\end{eqnarray}

and, from Eqs.~(\ref{eq:07}),~(\ref{eq:08}),  one gets the formula

\begin{eqnarray}
& &L_{e-e}^2 =\, (N - 1)d^2 \biggl(\frac{2l_p}{d} - 1\biggr) \, \nonumber
\\
& &l_p=\, \frac{1}{N-1} \sum_{i=1}^{N-1}l_p(i) \, ,
\label{eq:09}
\end{eqnarray}

relating the mean-square end-to-end separation to the average over all the local persistence lengths. Hence $l_p$, which measures the polymer stiffness, is determined  through the statistics of $L_{e-e}^2$.
Eq.~(\ref{eq:09}) holds for any linear chain with orientational correlation and reduces to the freely jointed chain result when $l_p(i) =\,d$ \cite{nelson03}.  Then, the chain with discrete bonds has an intrinsic stiffness, even in the absence of stacking interactions. 

Moreover, note that:

\textit{a)} Our definition refers to the total persistence length whereby we do not separate the electrostatic contribution from the intrinsic one as instead it is done in models which treat DNA as a charged polyelectrolyte and focus on the salt dependence of the persistence length \cite{fix,odi,weill,cherstvy,hsiao,manghi15}. Nevertheless our model can account for Coulomb repulsions due to negatively charged strands through the effective Morse potential parameters and for the screening effects arising from metal ions in solution through the solvent potential parameters as described in Section III.

\textit{b)} $l_p(i)$ is written in terms of the vectors $\textbf{d}_i$ connecting the centers of adjacent base pairs. However it may be also defined using the vectors normal to the base pair planes \cite{ejte}. As the $\textbf{d}_i$ are taken normal to the base pair planes (see Fig.~\ref{fig:1}(b)), any ambiguity regarding the definition of $l_p(i)$ is ruled out in this regard. 

\textit{c)} If the molecule is stretched, i.e. $L_{e-e}=\,L$, then $l_p$ attains the maximum, $l_p=\, N d/2$, allowed by the definition in Eq.~(\ref{eq:09}).

\textit{d)} While $l_p$ is an average measuring the global flexibility of the chain, the local flexibility of the DNA ends may generally differ from that of the central base pairs. In fact, alternative estimates of the persistence length can be made \cite{tan13} taking $l_p$ as the largest single contribution to Eq.~(\ref{eq:09}), i.e., $l_p \equiv l_p(i=1)$ \, which would lead to significantly higher values.

The sum in Eq.~(\ref{eq:08}) is truncated at the chain end no matter whether the correlation function has decayed to zero or not. In general end effects are less important the longer the molecule then, in chains with large $N$, one expects $< \cos \phi_{i,j} > \sim 0$ for most sites before $j$ gets of order $N$. Accordingly, one posits an exponential decay for the correlation function,  $< \cos \phi_{i,j} > =\,\exp \bigl(-\frac{|i - j|d}{A} \bigr) $ where the characteristic length scale $A$ may be in principle different from the microscopic definition for $l_p$ in Eqs.~(\ref{eq:08}),~(\ref{eq:09}).

The same idea of an exponential decay underlies the WLC model which provides a coarse grained description for the polymer as a continuous chain \cite{cifra}.
Assuming that the contour length $L$ remains finite  whereas $N \rightarrow \infty $, then $d$ becomes infinitesimally small and  
the sums in Eq.~(\ref{eq:07}) transform into integrals. Accordingly $L_{e-e}^2$ becomes:

\begin{eqnarray}
L_{e-e}^2 =\,  \int_{0}^{L}ds \int_{0}^{L}ds' \exp \biggl(-\frac{|s - s'|}{l_p^{WLC}} \biggr) \, ,
\label{eq:10}
\end{eqnarray}

where $s$, $s'$ are the arc length variables along the inextensible chain and $l_p^{WLC}$ is the WLC persistence length \cite{kp}. 
After computing the end-to-end-distances with Eqs.~(\ref{eq:06}),~(\ref{eq:07}), we compare the predicted values for 
$l_p$ via Eq.~(\ref{eq:09}) and $l_p^{WLC}$ via Eq.~(\ref{eq:10}). This permits to check whether and to which extent the $L_{e-e}$'s of the discrete model are consistent with the WLC formula. It is understood that, fitting by Eq.~(\ref{eq:10}) the $L_{e-e}$'s of the discrete model for short chains, one may obtain $l_p^{WLC}$'s at odds with the WLC predictions. 
In fact, strictly speaking, the WLC model is applicable under the assumption $l_p^{WLC} \ll L$ which may not be fulfilled for the short chains here considered. Furthermore,  Eq.~(\ref{eq:10}) in itself sets no upper bound in the range of possible values for $l_p^{WLC}$ which in principle may get larger than the $l_p$ estimated by Eq.~(\ref{eq:09}). This important difference between microscopic and continuous models will become evident at the light of the results presented hereafter.

Alternatively $l_p^{WLC}$ can be determined, as shown below, by relating it to the radius of gyration  \cite{doty,reed} which measures the global size of a polymer and can be experimentally assessed by small-angle x-ray scattering.

\section*{VI. Results and Discussion}

\begin{figure}
\includegraphics[height=8.0cm,width=8.0cm,angle=-90]{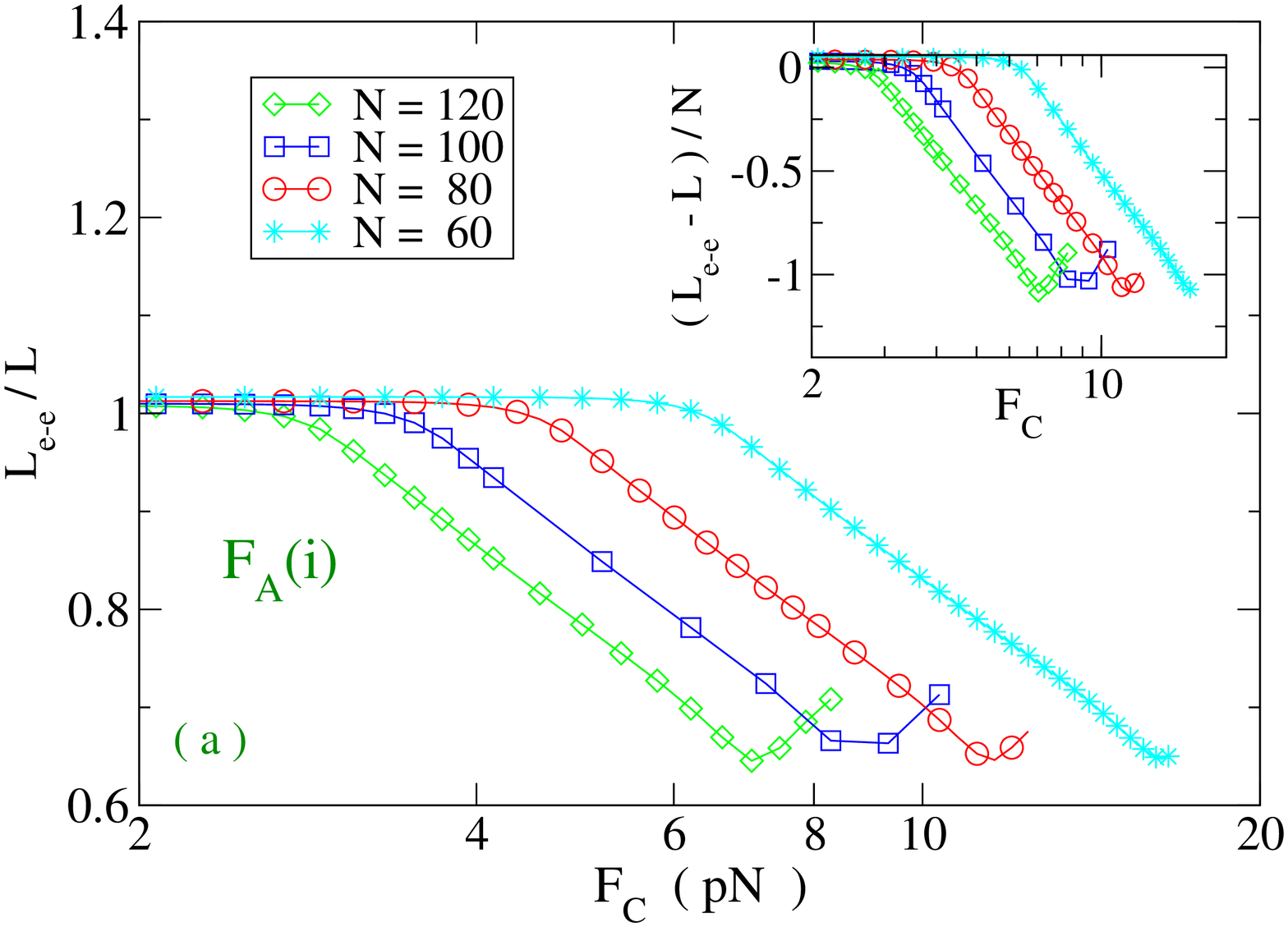}
\includegraphics[height=8.0cm,width=8.0cm,angle=-90]{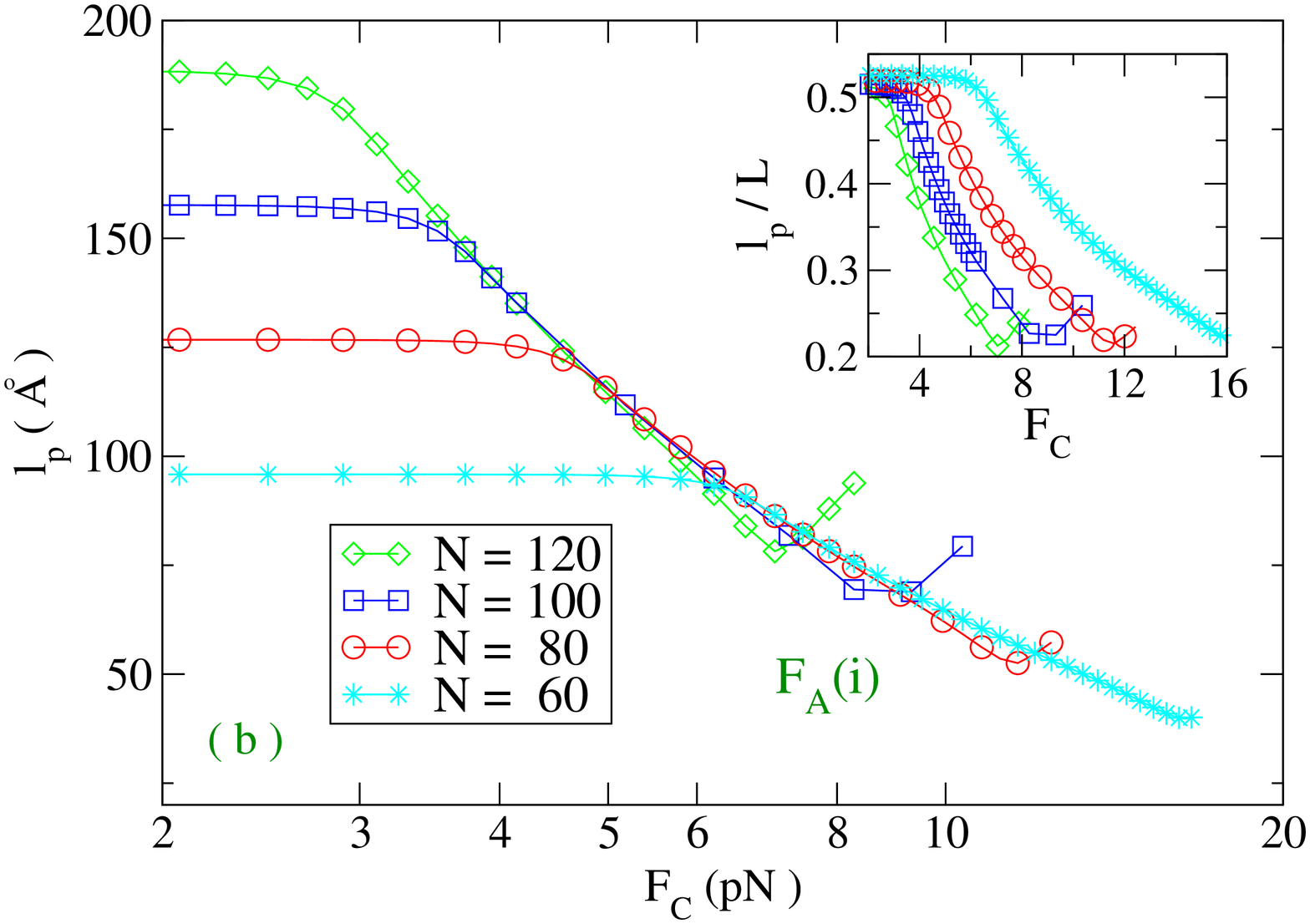}
\caption{\label{fig:3}(Color online)  
(a) {} End-to-end distances as function of the force in Eq.~(\ref{eq:07a}) with $F_C$ in the pico-Newton range. Four short chains are taken. The $L_{e-e}$ values are calculated from  Eq.~(\ref{eq:06}),~(\ref{eq:06a}),~(\ref{eq:07}) and are normalized over the contour length $L$. The inset shows the chain length contraction (in units \AA) per base pair.  (b) {} Persistence lengths corresponding to the chains in (a) and calculated from Eq.~(\ref{eq:09}). In the inset, the persistence lengths are normalized over $L$. }
\end{figure}

First we test our computational method by simulating a force profile $F_A(i)$, see Fig.~\ref{fig:3}, which linearly increases from the center to the chain ends:

\begin{eqnarray}
F_A(i)  =\,
\left\{\begin{matrix}
& &  2 \,F_{C} \Bigl(\frac{i - N/2}{N}      \Bigr)    \hskip 3cm   i \geq  N/2  \\ 
& &    \,F_{C} \Bigl(\frac{N/2 - i }{N/2 -1}\Bigr)    \hskip 3cm   i \leq  N/2
\end{matrix} \right.
\label{eq:07a}
\end{eqnarray}

The ratio $L_{e-e} / L$ is plotted in Fig.~\ref{fig:3}(a) for four homogeneous chains with $N=\,60,\,80,\,100,\,120$ \, as a function of $F_C$, i.e., the maximum  $F_A(i)$ at the chain ends. Also the end-to-end contractions per base pair are plotted in the inset.
All chains are initially aligned along the z-axis in the absence of forces.  Weak applied forces cannot balance the disordering action of contractile forces, accordingly, \, $L_{e-e} / L < 1$  in the weak forces regime. Instead, when a sufficiently high $F_C$ is applied, a distinctive upturn shows up and the ratio begins to grow versus $F_C$. Some striking differences among the chains appear: \textit{1)} the end-to-end contraction is more pronounced for larger $N$ consistently with the expectation that entropic forces are more effective in longer sequences. \textit{2)} the upturn shifts at larger $F_C$ by decreasing $N$: shorter  chains can also be bent by thermal fluctuations but larger forces are required to straighten the bonds due to the fact that the stiffness increases by decreasing $N$.

The average persistence lengths  for the profile $F_A(i)$ are shown in Fig.~\ref{fig:3}(b) together with the normalized values in the inset. While shorter chains have lower $l_p$ in the weak forces regime consistently with the definitions in Eqs.~(\ref{eq:08}),~(\ref{eq:09}), all $l_p$'s become comparable within a window of forces and eventually grow once the applied forces are large enough to straighten the chains. Interestingly, a somewhat different trend is found for the normalized $l_p$'s which display larger values for shorter chains in the weak forces regime, signaling that an intrinsic stiffness exists at short length scales in accordance with the results in Fig.~\ref{fig:3}(a).

\begin{figure}
\includegraphics[height=8.0cm,width=8.0cm,angle=-90]{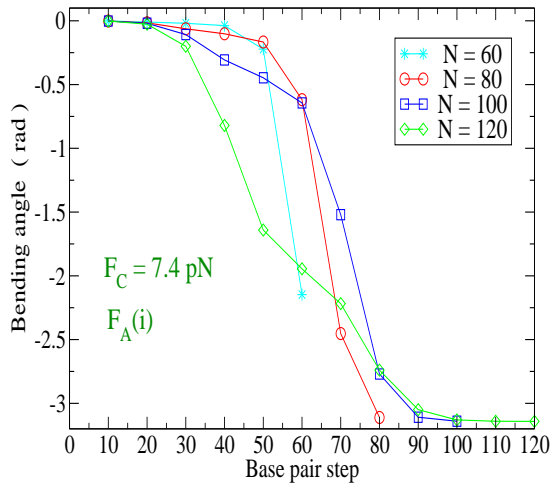}
\caption{\label{fig:4}(Color online)  
Average bending angles (formed by the segments $d_i$ with respect to the z-axis in Figs.~\ref{fig:2}) versus the base pair site. The force profile in Eq.~(\ref{eq:07a}) with a sizeable value for $F_C$ is applied to the four chains.}
\end{figure}

Also the average bending angles, obtained from Eq.~(\ref{eq:06}) with the applied force $F_A(i)$, are displayed in Fig.~\ref{fig:4} for the four chains. Precisely, the cumulative average angle with respect to the z-axis (marked by the red arrows in Figs.~\ref{fig:2})) is plotted versus the base pair site. 
The chosen $F_C=\,7.4 \,pN$ corresponds to the minimum $L_{e-e} / L$ for the chain with $N=\,120$ in Fig.~\ref{fig:3}(a). For all other chains, that value belongs to the range of weak applied forces in which the entropic forces are dominant.    
The $N=\,120$ chain is substantially bent by the contractile forces as shown in Fig.~\ref{fig:3}(a). Accordingly, we find that the last $20$ \textit{bps} bonds along the chain make an average angle $\sim \pi$ with respect to the first bond. By reducing the chain length, the number of \textit{bps} bonds bent by large angles becomes smaller and, for the $N=\,80$ chain, only the last bond  is bent  by $\pi$ with respect to the z-axis, namely the case depicted in Fig.~\ref{fig:2}(a).  As for the $N=\,60$ chain, although no base pair bond attains the maximum angle,  most of the bending is ascribed to the last ten bonds along the sequence, i.e. between the \textit{50}-th and\textit{ 60}-th base pair step. Then we find that, although the chain bendability decreases below $N \sim \,100$, shorter DNA chains still maintain a considerable flexibility with the stronger bending stemming from the unconstrained chain end of our model. This appears consistent with the results of molecular dynamics and Monte-Carlo simulations for a set of very short sequences \cite{tan} pointing to an enhanced flexibility due to six base pairs at each chain end. 

\begin{figure}
\includegraphics[height=8.0cm,width=8.0cm,angle=-90]{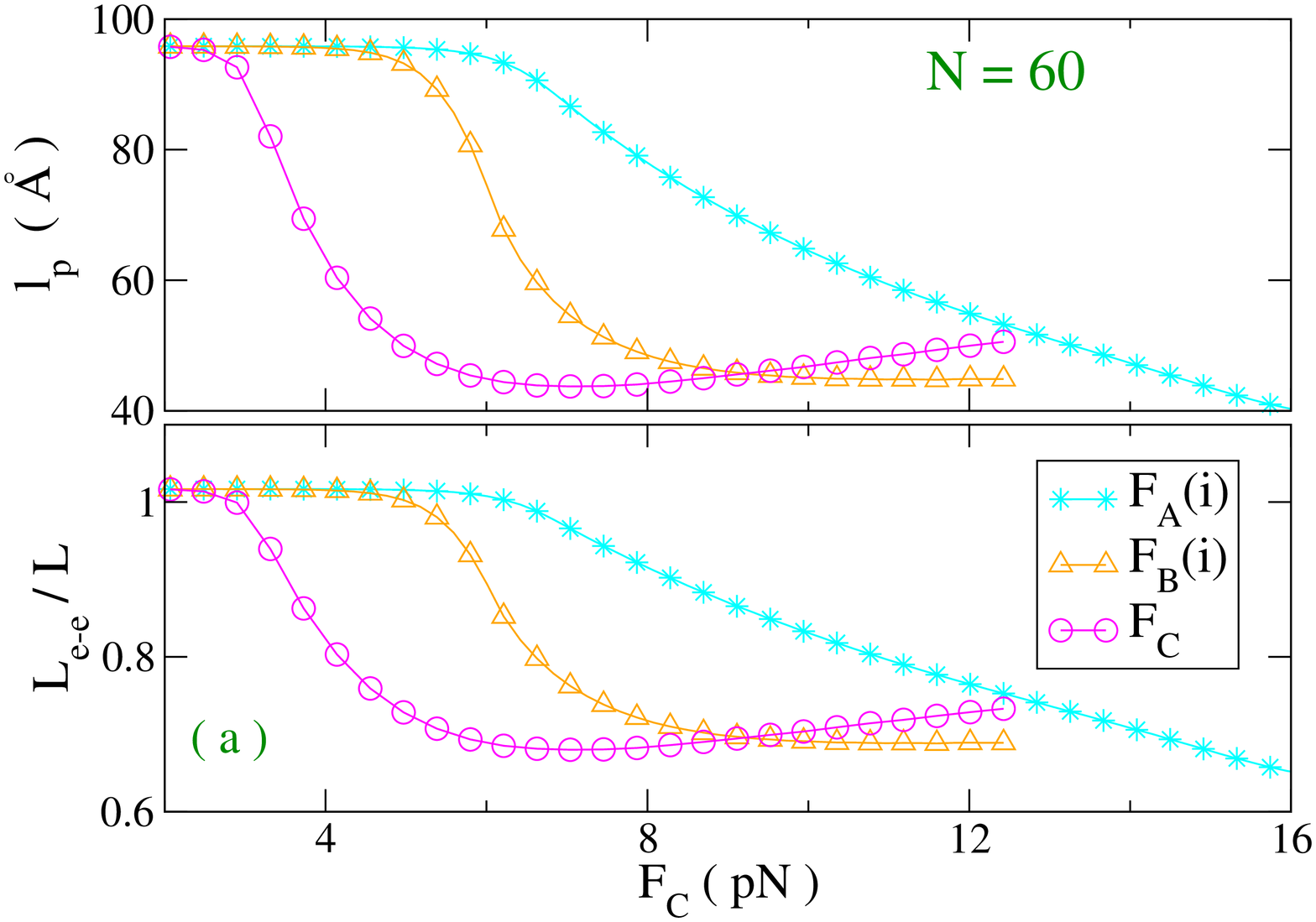}
\includegraphics[height=8.0cm,width=8.0cm,angle=-90]{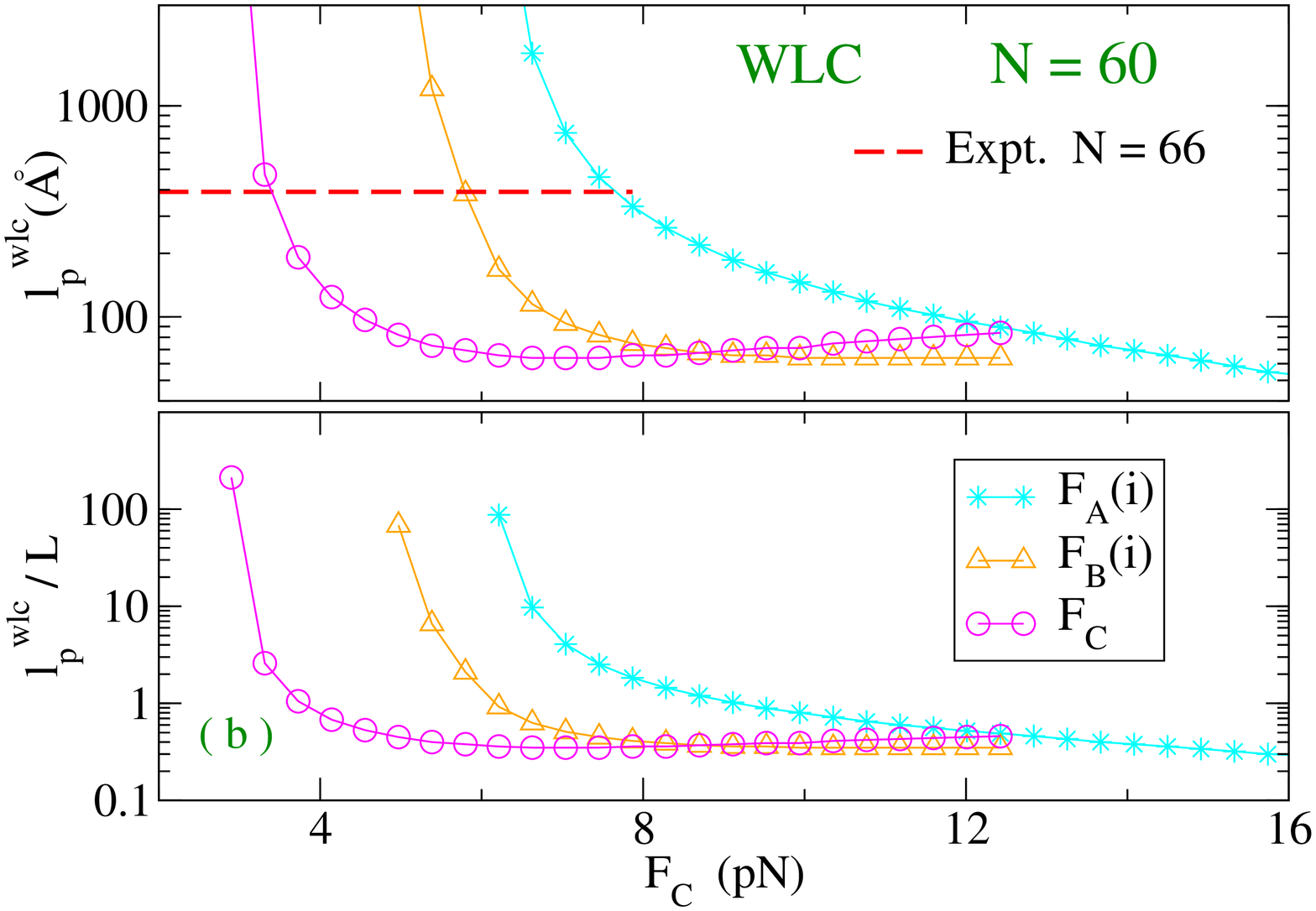}
\caption{\label{fig:5}(Color online)  (a) {}  Persistence lengths (upper panel) and end-to-end distances (lower panel) calculated from Eq.~(\ref{eq:09}) for the chain with $60$ base pairs. Three profiles of forces are considered. 
(b) {} Persistence lengths from the WLC formula in Eq.~(\ref{eq:10}) (with $L_{e-e}$'s values shown in (a)) for the same chain and forces as in (a). The dashed line represents the persistence length obtained by fitting the experimental radius of gyration for a $N=\,66$ chain given in \cite{archer}. The persistence lengths, normalized over the chain length, are plotted in the lower panel.}
\end{figure}

Next we check to which extent the analysis presented so far depends on the specific choice of the applied force $F_A(i)$. Alternatively, we take a force $F_B(i)$ whose intensity decreases from the maximum $F_C$ at the chain center to the zero at the chain ends, i.e.:

\begin{eqnarray}
F_B(i)  =\,
\left\{\begin{matrix}
& &  2 \,F_{C} \Bigl( \frac{N - i}{N}       \Bigr) \hskip 3cm   i \geq  N/2  \\ 
& &    \,F_{C} \Bigl( \frac{i - 1 }{N/2 -1} \Bigr) \hskip 3cm   i \leq  N/2
\end{matrix} \right.
\label{eq:07b}
\end{eqnarray}

and a force which is constant throughout the chain with value $F_C$. This latter case is the one usually assumed to model the force-extension behavior and interpret the experiments. We focus on the shortest among the chains of our set. For the $N=\,60$ chain, the $l_p$ and $L_{e-e}$'s obtained from Eq.~(\ref{eq:09}) with the profiles  $F_A(i)$, $F_B(i)$ and constant $F_C$ are compared in Fig.~\ref{fig:5}(a). Essentially it is found that: 

\textit{1)} weak applied forces induce the molecule elastic response which, however, strongly depends on the force intensity at the specific base pair sites. In fact, relatively larger forces applied at the chain center ($F_B(i)$), reduce $l_p$ with respect to $F_A(i)$. This is explained by noticing that the segments at the chain ends have anyway less spatial constraints than those in the center, for any applied force. Hence $l_p$ is more significantly reduced and the global flexibility is increased by perturbing the chain center, that makes the stiffer part of the molecule. 

\textit{2)} Remaining in the weak forces regime, i.e. below $ \sim 7 \,pN$, a constant $F_C$ along the chain further reduces $l_p$ which quickly drops from the $Nd/ 2$ value associated to the fully stretched conformation. However, above $ \sim 7 \,pN$, the applied force is strong enough to stretch the chain and $l_p$ grows. In this intermediate force regime, $l_p$ grows more quickly under the effect of a uniform $F_C$.  Instead, for the profiles $F_B(i)$ and $F_A(i)$,  the molecule stretching begins at respectively larger forces. Thus, the transition between weak forces regime (with decreasing $l_p$) and intermediate forces regime (with increasing $l_p$) depends  on the specific perturbations applied to the base pair sites of the short chain. 

Certainly, the reported $F_C$'s should be taken only as indicative of a qualitative behavior whereas their effective values may vary with the model input parameters chosen to model specific sequences.
With this caveat, we plot in
Fig.~\ref{fig:5}(b) both the $l_p^{WLC}$ (upper panel) and the normalized values (lower panel) obtained from Eq.~(\ref{eq:10}), by fitting the calculated $L_{e-e}$ for the same chain and force profiles considered in Fig.~\ref{fig:5}(a). Consistently with the definition in the WLC model \cite{volo05},  $l_p^{WLC}$ tends to very large values for a  stretched conformation i.e., for  $L_{e-e} \sim  L$. This marks a main difference between the continuous model and the discrete model of Eq.~(\ref{eq:09}). 
For the coiled conformations with weak applied forces, $l_p^{WLC}$ quickly drops but remains much larger than $l_p$. However, by increasing $F_C$, $l_p^{WLC}$ and $l_p$ get closer for all types of forces. For instance, taking a constant profile with $F_C \sim 10 \,pN$, we find $l_p^{WLC} \sim 72 \AA$ whereas, from Fig.~\ref{fig:5}(a),  $l_p \sim 47 \AA$. Furthermore, for the constant force profile, it is found that  the condition \, $l_p^{WLC} < L$ \, (see lower panel) is  fulfilled for $F_C > 3.8 \, pN$ . 
Altogether, we conclude that the use of Eq.~(\ref{eq:10}) overestimates $l_p^{WLC}$ as already noted also for longer sequences \cite{manghi15}. Moreover, the $l_p^{WLC}$'s here derived show a significant variation with $F_C$ at variance with the WLC model itself whose persistence length is almost independent of the applied force \cite{croque}.

Also the persistence length derived from the measured radius of gyration, $R_g \sim 60 \AA$, for a $N=\,66$ chain \cite{archer}, is shown in  Fig.~\ref{fig:5}(b) (red dashed line).
In order to fit  $R_g$, we use the WLC relation \cite{doniach}, \, $(R_g^{WLC})^2 / L^2=\, x/3 - x^2 + 2x^3 - 2x^4 \bigl(1 - \exp(-1/x) \bigr) $ with $x \equiv l_p^{exp} / L$ \,  and $l_p^{exp}$  denotes the value obtained from the experimental $R_g$. Precisely, $R_g$ is corrected by involving the finite DNA diameter $R_0$, i.e.,  $ R_g^2=\, (R_g^{WLC})^2 + R_0^2 /8$.   Taking $R_0=\,20 \AA$ as above, we get $l_p^{exp} \sim 391 \AA$  which is lower than the length ($450 \AA$) obtained in ref.\cite{tan} from the experimental $R_g$ of ref.\cite{archer}.  It is however found that $l_p^{exp}$ is very sensitive to the input parameters used in the fitting. On the other hand, $l_p^{exp}$ remains well above the range of values predicted for the $N=\,60$ chain in Fig.~\ref{fig:5}(a) by the discrete model. This confirms quantitatively that the latter leads to more conservative estimates of the persistence length than those made using the WLC model. It is also noticed that $l_p^{exp}$ is much larger than $L$ and this questions the applicability of the WLC  formula \cite{busta94}  to force-extension experiments in short chains.

The dashed line intersects the plots derived from Eq.~(\ref{eq:10}) at distinct force values in the weak regime for the three profiles. Thus, at the intersection points, $l_p^{exp}$ matches the $l_p^{WLC}$'s obtained by fitting the computed $L_{e-e}$. This feature has interesting potential applications.
For instance, assuming that the $l_p^{exp}$'s for the $N=\,66$ and $N=\,60$ chains are similar and assuming e.g., a uniform external force with $F_C \sim 3.5\, pN$, one expects from Fig.~\ref{fig:5}(a) (lower panel) that the molecule is in a bent conformation with $L_{e-e} \sim 0.9 L$. Although at this stage there are not experimental informations to check such predictions, we propose that this approach based on independent estimates of $l_p^{exp}$ and $L_{e-e}$ can provide consistent indications to further elucidate the issue of the validity of the WLC model in very short chains.

\begin{figure}
\includegraphics[height=8.0cm,width=8.0cm,angle=-90]{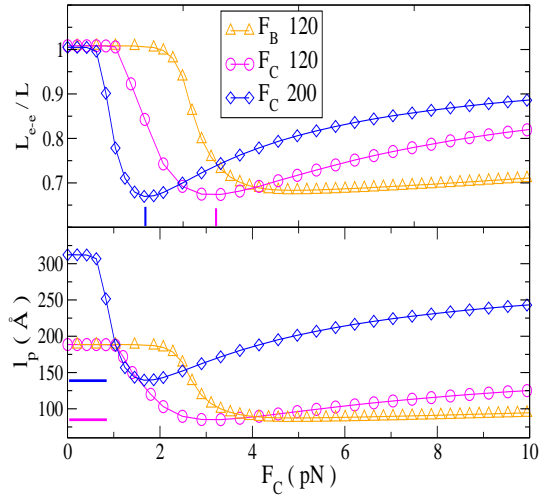}
\caption{\label{fig:6}(Color online)  Normalized end-to-end distances (upper panel) and persistence lengths (lower panel) calculated from Eq.~(\ref{eq:09}) for the chain with $120$ base pairs (with $F_B(i)$ and uniform $F_C$ forces) and $200$ base pairs (with $F_C$).  The minimum values, which denote the most bent molecules configurations in the presence of uniform $F_C$, are marked in both panels.
}
\end{figure}

By increasing $N$, the transition between decreasing and increasing $l_p$'s shifts at lower forces as already pointed out in Fig.~\ref{fig:3}(b) for the profile $F_A(i)$. The trend, albeit more pronounced, is found for the profile $F_B(i)$ and mostly for a uniform $F_C$ as shown, for $N=\,120$  in Fig.~\ref{fig:6} where the normalized end-to-end distances and persistence lengths are drawn. Also the plots for the $N=\,200$ chain with uniform $F_C$ are reported.

The force-extension behavior with $L_{e-e}$ growing as a function of the applied force, is recovered from $F_C \sim 3.3 pN$ ($N=\,120$) and $F_C \sim 1.6 pN$ ($N=\,200$) upwards. It is thus expected that for longer chains (in the kilo-\textit{bps} range) than those considered here, the usual mechanical response of the entropic elastic regime begins at  weaker stretching perturbations and  also $F_C$ of order \, $\sim 0.1 \,pN$ or lower can produce some extension of the molecule. 

Eventually, noticing that the persistence length defined by the discrete model varies with the applied force, the question arises as to weather such quantity can provide a realistic measure of the overall stiffness of the chain. On the base of the presented calculations, we propose that a stiffness indicator can be extracted from the minimum values in the $l_p$'s plots, marked in Fig.~\ref{fig:6} for two chains. In fact, the minima correspond to the most bent chains configuration under the contractile effect of the entropic forces, as visualized in the upper panel.

\section*{VII. Conclusions }

We have studied the flexibility of short chains with $N$ base pairs by computing the average bending angles between adjacent base pairs along the molecule backbone. The calculation is based on a  mesoscopic Hamiltonian that models both the inter-strand forces with hydrogen bonds between the base pair mates and the intra-strand forces between stacked base pairs, also depending on the rotational degrees of freedom. Thus, experimentally accessible quantities such as end-to-end distance and persistence length are related to the effective parameters of the potential in a discrete model which treats the chain at the level of the base pair, it includes the effects of the solvent and allows for kinks formation due to large bending fluctuations between neighbor base pairs along the stack.
While the proposed approach is general enough to be applied to any heterogeneous sequence in solution, we have considered in this paper a set of short homogeneous fragments to focus on the interplay between chain bendability and its length.

The ensemble averages for the bending angles are performed by an integration over a large distribution of base pair configurations which guarantees the thermodynamic equilibrium of the system and accounts for the thermal fluctuations of the base pair separations. This feature also permits to investigate the chain flexibility as a function of temperature, a subject which is left for a future work. Here we have studied, at room temperature, the chain elastic response to an applied mechanical stretching represented both by forces whose intensity varies along the chain and by a constant force field as usually taken in force-extension analysis. Importantly, the study has been limited to a regime of weak and intermediate forces on the pico-Newton scale consistent with the assumption that rise distance and structure of the molecule are not deformed. 
Moreover, our calculation applies to torsionally constrained chains as the helical repeat of the molecule, i.e. the twist angle, is taken as constant. This constraint may be removed in a separate study by determining, for any applied force, the specific twist conformation which minimizes the free energy of the chain, albeit at the expense of a much longer computational time.

Assuming that in the absence of external forces the short molecule is stretched, i.e., the end-to-end distance is equal to the contour length,  we have shown that under the action of a weak perturbation the chain response is dominated by thermal bending fluctuations which contract $L_{e-e}$ down to a minimum value. By enhancing the applied force over a certain threshold, the bonds along the molecule backbone are progressively straightened and $L_{e-e}$ grows as a function of the force. Interestingly, the value of the threshold depends on the type of applied force and, for a specific profile e.g., $F_A(i)$ in Fig.~\ref{fig:3}(a) or a uniform force in Fig.~\ref{fig:6},  the  threshold shifts upwards by decreasing the chain length. This is physically understood by observing that, once the molecules are in the entropically favored bent conformation, larger forces are required to align the bonds of shorter chains which are intrinsically stiffer. This interpretation may seem at variance with the 
displayed persistence lengths which are indeed higher for longer chains. However, the apparent contradiction is solved upon noticing that the calculated $l_p$'s are averaged over the site dependent persistence lengths of our discrete model and, for the stretched configuration without applied forces, are proportional to $N$. Once the $l_p$'s are normalized over the contour length $L$, the trend is in fact reversed and the ratios $l_p/ L$ of the bent conformations are higher for shorter chains (see inset in Fig.~\ref{fig:3}(b)) confirming that the latter have an enhanced stiffness \textit{per base pair}. Notwithstanding, the predicted $l_p$'s are significantly smaller than those estimated for long DNA molecules through the worm-like-chain model and even smaller than the value experimentally obtained for a short chain with comparable number of base pairs. In this regard we have discussed the possible sources of these discrepancies and pointed out that the WLC formulas, relating the persistence length to the end-to-end distance and to the radius of gyration respectively, yield overestimated values with respect to the discrete model. 
The fact remains that several ways to estimate the persistence length appear in the literature and this contributes to contrasting outcomes. 

Altogether our results are in line with a body of recent studies which have emphasized the remarkable flexibility of DNA on short length scales although more extensive simulations and stricter comparisons with experiments on the same fragments are necessary to elucidate these questions.

\end{document}